\documentclass[conference]{IEEEtran}
\usepackage{amsfonts}
\IEEEoverridecommandlockouts

\ifCLASSINFOpdf
\else
\fi
\usepackage{epsfig}
\usepackage{graphicx}
\usepackage{psfig}
\usepackage{epsf}
\usepackage[cmex10]{amsmath}
\usepackage{booktabs}
\usepackage{fancyhdr}
\usepackage{slashbox}
\usepackage{caption2}   

\newtheorem{lemma}{\textbf{Lemma}}
\begin{document}
\title{Resource Allocation Based on Deep Neural Networks for Cognitive Radio Networks}
\author{Fuhui Zhou$^{\dag}$$^{\S}$, Xiongjian Zhang$^{\dag}$, Rose Qingyang Hu$^{\S}$, Apostolos Papathanassiou$^{\ddag}$, Weixiao Meng$^{\pounds}$\\
$^{\S}$Utah State University, USA, $^{\dag}$Nanchang University, China,  $^{\pounds}$Harbin Institute of Technology, China\\
$^{\ddag}$Intel Corporation \& Intel Platform Engineering Group, USA \\

Email: \emph{\{zhoufuhui@ieee.org, rose.hu@usu.edu, Apostolos.Papathanassiou@intel.com, wxmeng@hit.edu.cn\}}
\thanks{The research was supported by the National Science Foundation under the grants NeTS 1423348 and EARS-1547312, the National Natural Science Foundation of China under the Grant 61701214 and Grant 61728104, in part by the Young Natural Science Foundation of Jiangxi Province under Grant 20171BAB212002, in part by The Open Foundation of The State Key Laboratory of Integrated Services Networks under Grant ISN19-08, and in part by The Postdoctoral Science Foundation of Jiangxi Province under Grant 2017M610400, Grant 2017KY04 and Grant 2017RC17.}}

\maketitle
\begin{abstract}
Resource allocation is of great importance in the next generation wireless communication systems, especially for cognitive radio networks (CRNs). Many resource allocation strategies have been proposed to optimize the performance of CRNs. However, it is challenging to implement these strategies and achieve real-time performance in wireless systems since most of them need accurate and timely channel state information and/or other network statistics. In this paper a resource allocation strategy based on deep neural networks (DNN) is proposed and the training method is presented to train the neural networks. Simulation results show that our proposed strategy based on DNN is efficient in terms of the computation time compared with the conventional resource allocation schemes.
\end{abstract}
\IEEEpeerreviewmaketitle
\section{Introduction}
\IEEEPARstart{T}{he} spectrum scarcity problem is increasingly  severe due to the unprecedented proliferation
of mobile devices and the rapidly growing demand on the broadband communication services \cite{F. Zhou4}. In order to alleviate this problem, cognitive radio (CR) has been proposed and received considerable attention in the past decade  \cite{Tragos}. In CR networks (CRNs), the secondary users (SUs) can access the spectrum bands of the primary users (PUs) on the conditions that the interference caused to the PUs is tolerable \cite{F. H. Zhou1}. There are three operational modes in CRNs, namely, interweave, overlay, and underlay \cite{F. Zhou4}-\cite{F. H. Zhou1}. Under the interweave mode, spectrum sensing is required to find the available spectrum bands. Under the overlay mode, the SUs cooperate with the PUs and obtain the opportunity to access the spectrum bands of the PUs. For the underlay mode, the SUs can coexist with the PUs provided that the interference caused to the PUs is tolerable. In this paper, we focus on the underlay mode due to its high spectral efficiency (SE) and ease of implementation \cite{F. Zhou4}-\cite{F. H. Zhou1}.

Meanwhile, how to improve energy efficiency (EE) is of vital importance  in the next generation wireless communication systems since it can decrease the greenhouse gas emission and achieve a sustainable operation \cite{F. Zhou3}.  It has been reported in \cite{F. H. Zhou1}, \cite{F. Zhou3} that  2$\%$ of the greenhouse gas and 2$\%$ to 10$\%$ of global energy consumption are caused by information and communication technologies. Thus, it is important to develop energy-efficient transmission techniques for the wireless communication systems, especially for CRNs. The reason is that energy-efficient transmission techniques can not only prolong the operational time of SUs, but also well protect the PUs from the harmful interference.

There have been many research works devoted to designing resource allocation strategies for improving SE and EE of CRNs \cite{F. H. Zhou1}, \cite{X. Kang}-\cite{A. Yaqot}. The SE of CRNs was maximized by the proposed resource allocation strategies under the average/peak power constraint in \cite{X. Kang}. In \cite{X. Kang1}, the authors extended the SE maximization problem into the outage probability constraint and proposed the optimal resource allocation strategies. In contrast to the work in \cite{X. Kang1}, the authors in \cite{F. H. Zhou1} designed energy-efficient resource allocation strategies for maximizing the achievable EE of the secondary network. It was shown that there is a tradeoff between EE and SE in CRNs. In \cite{A. Alabbasi}, the authors have extended the EE maximization problem into CRNs with the opportunistic spectrum access mode. The sensing time and the transmission power of the cognitive base station (CBS) were jointly optimized to maximize the EE of the SUs. In order to further improve the EE of the secondary network, multiple-input multiple-output (MIMO) techniques were exploited in CRNs and the energy-efficient precoding scheme was designed \cite{X. Zhang}. In \cite{F. H. Zhou1}, \cite{X. Kang1}-\cite{X. Zhang}, the average/peak interference power constraint was applied to protect the quality of service (QoS) of the PUs. Different from the works in \cite{F. H. Zhou1}, \cite{X. Kang1}-\cite{X. Zhang}, the authors in \cite{L. Wang} leveraged the outage probability constraint to protect the QoS of the PUs. It was shown that the EE can be further improved by using this constraint. Recently, in order to improve the SE of the secondary network,  MIMO techniques and orthogonal frequency-division multiplexing (OFDM) techniques were exploited in CRNs and the optimal resource allocation strategy was designed \cite{A. Yaqot}.

Note that the resource allocation strategies proposed in \cite{F. H. Zhou1}, \cite{X. Kang}-\cite{A. Yaqot} were obtained by using iterative or alternative algorithms, irrespective of  SE or EE maximization. Moreover, the sub-gradient algorithm was applied to update the dual variables. The implementation complexity increases with the number of users in CRNs, which makes  it challenging to apply these schemes  in the Internet of Things with  massive number of users in the networks. What's worse, these resource allocation strategies need the perfect channel state information and the accurate information about the CRNs, which are normally very difficult to acquire in practice, especially when massive users exist.

Motivated by the fact that the exploitation of machine learning to design resource allocation schemes has tremendous potentials to reduce the implementation complexity and to achieve real-time performance \cite{M. Chen1}-\cite{F. Zhou5}, the  deep neural networks (DNN) are applied to develop resource allocation strategies for maximizing EE and SE of CRNs. A resource allocation framework based on DNN is proposed to optimize  SE and  EE of CRNs. The training method is given to obtain the parameters of the DNN. Simulation results show that our proposed resource allocation strategies based on DNN are efficient in terms of the real-time implementation and the system performance.

The remainder of this paper is organized as follows.  Section II illustrates the system model. The resource allocation problems are examined in Section III. Section IV presents simulation results. The paper concludes with Section V.
\section{System Model}
\begin{figure}[!t]
\centering
\includegraphics[width=3.3 in]{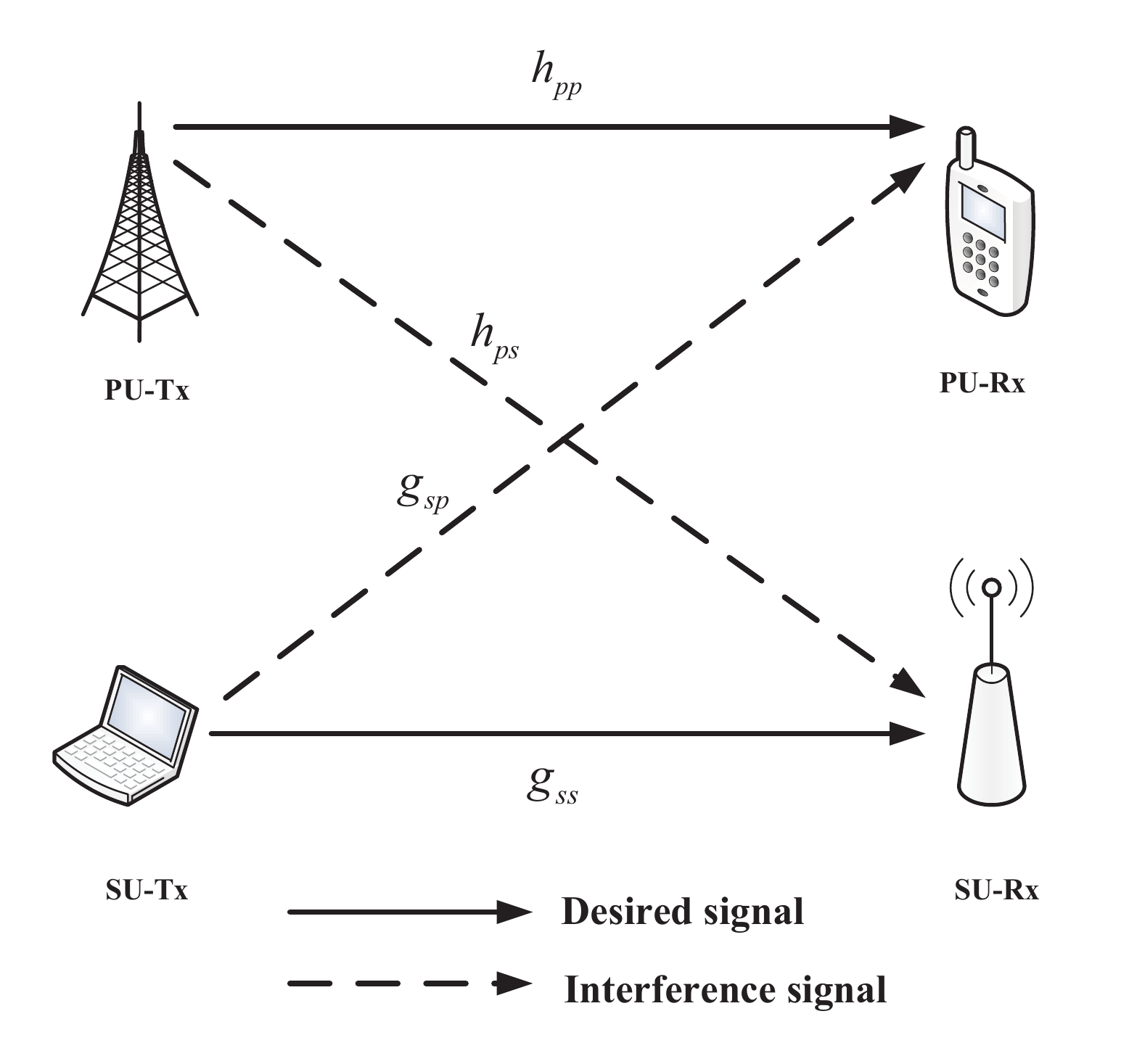}
\caption{The system model.} \label{fig.1}
\end{figure}
As shown in Fig. 1, similar to \cite{F. H. Zhou1}, \cite{X. Kang1}, and  \cite{L. Wang}, a simplified CRN is considered in order to permit meaningful insights into the design of resource allocation strategies based on DNN. In the CRN, the secondary network coexists with the primary network by using the underlay mode. In this case, the PBS and the CBS can transmit signals  on the same frequency channel. The primary link consists of one primary base station (PBS) and one primary user (PU) while the secondary link has one CBS and one secondary user. All the terminals are equipped with one antenna. Similar to \cite{F. H. Zhou1}, \cite{X. Kang1},  and  \cite{L. Wang}, it is assumed that all the channels are  block fading channel.  All the channel power gains are independent identically distributed (i.i.d.) ergodic and stationary random variables with continuous probability density functions.

In order to protect the QoS of the PU, an interference power constraint is imposed. According to \cite{F. H. Zhou1} and \cite{X. Kang1}, compared with the peak interference power constraint, the average interference power (AIP) constraint can not only well protect the QoS of the PU, but also improve the performance of the SU. Thus, the AIP constraint is applied, given as
\begin{align}
\mathbb{E}\left\{ {{g_{sp}}{P_s}} \right\} \le {{P_{In}}},
\end{align}
where ${{P_{In}}}$ denotes the maximum AIP that can be tolerable for the PU; $P_s$ denotes the transmit power of the CBS; $\mathbb{E}\left( \cdot \right)$ represents the expectation operator, and $g_{sp}$ is the  instantaneous channel
power gain from the CBS to the PU. Moreover, in order to  satisfy the long-term power budget of the CBS, the average transmit power (ATP) constraint is considered, given as
\begin{align}
&\mathbb{E}\left\{ {{P_s}} \right\} \le {{P_{th}}} , {P_s} \ge 0,
\end{align}
where $ {{P_{th}}}$ denotes the maximum ATP of the SU.
\section{Resource Allocation Problems}
In this section, the  SE and EE maximization problems are studied. The resource allocation framework based on DNN is given. Moreover, the training method for the DNN is presented.
\subsection{Problem Formulations}
In order to maximize the average transmit rate of the SU, the SE maximization problem under the AIP and the ATP constraints can be formulated as $\text{P}_{\text{1}}$, given as
\begin{subequations}
\begin{align}\label{27}\
\text{P}_{\text{1}}: \ \ \ \ \ &{{\mathop {\max }\limits_{{P_s}} } }\ {{{{R_{SE}}\left( {{P_s}} \right)}} = {\mathbb{E}\left\{ {{{\log }_2}\left( {1 + \frac{{{g_{ss}}{P_s}}}{{{h_{ps}}{P_p} + \sigma _w^2}}} \right)} \right\}}}  \\
&\text{s.t.}\ \ \  \left(1 \right) \text{and} \left(2 \right) \text{are satisfied},
\end{align}
\end{subequations}
where $R _{SE}\left( {{P_s}} \right)$ denotes the average transmit rate of the SU; $g_{ss}$ and $h_{ps}$ are the instantaneous channel power gains of the secondary link and the link from the PBS to the SU; $P_p$ is the transmit power of the PBS; $\sigma _w^2$ is the variance of noise at the SU. In this paper, similar to \cite{F. H. Zhou1}, \cite{X. Kang1}, and  \cite{L. Wang}, the PBS employs a non-adaptive power transmission strategy and $P_p$ is a constant. Using the Lagrange duality method and the Lagrange dual-decomposition method \cite{X. Kang}, the authors have proposed the optimal power allocation strategy. For convenience, Lemma 1 is given to present this strategy.
\begin{lemma} \cite{X. Kang}
The optimal power allocation strategy for maximizing the SE of the CRNs under the AIP and ATP constraints is given by
\begin{align}\label{27}\
P_{se}^{opt} = {\left[ {\frac{1}{{\left( { \tau  + \mu {g_{sp}}} \right)\ln 2}} - \frac{{\left( {{h_{ps}}{P_p} + \sigma _w^2} \right)}}{{{g_{ss}}}}} \right]^ + },
\end{align}
where ${\left[ a \right]^ + } = \max \left( {a,0} \right)$ and $\max \left( {a,0} \right)$ represents the maximum between $a$ and $0$.  In $\left(4\right)$, $P_{se}^{opt}$ denotes the optimal transmit power for maximizing the SE of the CBS; $\tau$ and $\mu$ are the dual variables corresponding to the ATP and AIP constraints, respectively. These dual variables are obtained by using the sub-gradient method \cite{X. Kang}.
\end{lemma}

In contrast to the SE maximization problem, the EE maximization problem under the AIP and ATP constraints can be formulated in CRNs as
$\text{P}_{\text{2}}$, given as
\begin{subequations}
\begin{align}\label{27}\
\text{P}_{\text{2}}: \ \ \ \ \ &{{\mathop {\max }\limits_{{P_s}} } }\ {{{{\eta _{EE}}\left( {{P_s}} \right)}} = \frac{{\mathbb{E}\left\{ {{{\log }_2}\left( {1 + \frac{{{g_{ss}}{P_s}}}{{{h_{ps}}{P_p} + \sigma _w^2}}} \right)} \right\}}}{{\mathbb{E}\left\{ {\zeta {P_s}+ {P_C}} \right\}}}}  \\
&\text{s.t.}\ \ \  \left(3\rm{b} \right),
\end{align}
\end{subequations}
where $\eta _{EE}$ denotes the average transmit rate of the SU;  $\zeta$ and $P_C$ denote the amplifier coefficient and the constant circuit power consumption of the CBS, respectively. Based on the Dinkelbach's method and the Lagrange duality method,  in \cite{F. H. Zhou1}, we proposed an energy-efficient optimal power allocation algorithm to solve $\text{P}_{\text{2}}$ and derived the energy-efficient optimal power allocation strategy. Lemma 2 is given to present this energy-efficient resource allocation strategy.
\begin{lemma} \cite{F. H. Zhou1}
The energy-efficient optimal power allocation strategy for maximizing the EE of CRNs under the AIP and ATP constraints is given by
\begin{align}\label{27}\
P_{ee}^{opt} = {\left[ {\frac{1}{{\left( {\eta \zeta  + \tau  + \mu {g_{sp}}} \right)\ln 2}} - \frac{{\left( {{h_{ps}}{P_p} + \sigma _w^2} \right)}}{{{g_{ss}}}}} \right]^ + },
\end{align}
where $P_{ee}^{opt}$ denotes the energy-efficient optimal transmit power of the CBS; $\tau$ and $\mu$ are the dual variables corresponding to the ATP and AIP constraints, respectively. In $\left(6\right)$, $\eta$  is a non-negative cost factor and the dual variables are obtained by using the sub-gradient method \cite{F. H. Zhou1}.
\end{lemma}

Note that the sub-gradient method is required to obtain the optimal power allocation strategy for maximizing the SE of CRNs. Moreover, for the EE maximization problem, an iterative algorithm and the sub-gradient method are required to obtain the energy-efficient optimal power allocation strategy proposed in \cite{F. H. Zhou1}. The implementation complexity increases with the number of the users in CRNs. Furthermore, these resource allocation strategies are based on the perfect channel state information.  It is challenging to implement these strategies and achieve real-time performance in practice when there exist massive users.
\subsection{Resource Allocation Framework Based on DNN}
In this subsection, in order to obtain the  resource allocation strategy and achieve real-time implementation, a resource allocation framework based on DNN is proposed. The  training method for obtaining the characteristic parameters of the DNN is given. Let  $\mathcal{F}$ and $\mathcal{P}$ respectively denote  the feasible region and the parameter space of the formulated problem. Moreover, it is assumed that they are compact sets. The principle of this resource allocation framework is given by Lemma 3.
\begin{lemma} \cite{H. Sun}
The mapping from the problem parameter $z$ and the initial vaule $x_0$ to  the final output $x_f$ can be accurately estimated  by using the deep neural networks that have $N$ hidden units with the sigmoid activation function. Moreover, a large enough positive constant $N$ exists such that the upper bound of the error between the final output of the DNN and the desired value cannot be larger than the given error.
\end{lemma}

Based on Lemma 3, it is proved in \cite{S. Liang} that a deterministic algorithm that iteratively updates the continuous mappings can be learned by using a deep neural network. Since the optimal power allocation strategy for maximizing the SE of CRNs given by Lemma 1 is a continuous mapping between the channel power gains and the transmit power level of the CBS and each iteration of the energy-efficient optimal power allocation algorithm  proposed in \cite{F. H. Zhou1} is also a continuous mapping, a DNN can be exploited to learn the proposed power allocation strategies given by Lemma 1 and Lemma 2.
\begin{figure}[!t]
\centering
\includegraphics[width=3.6 in]{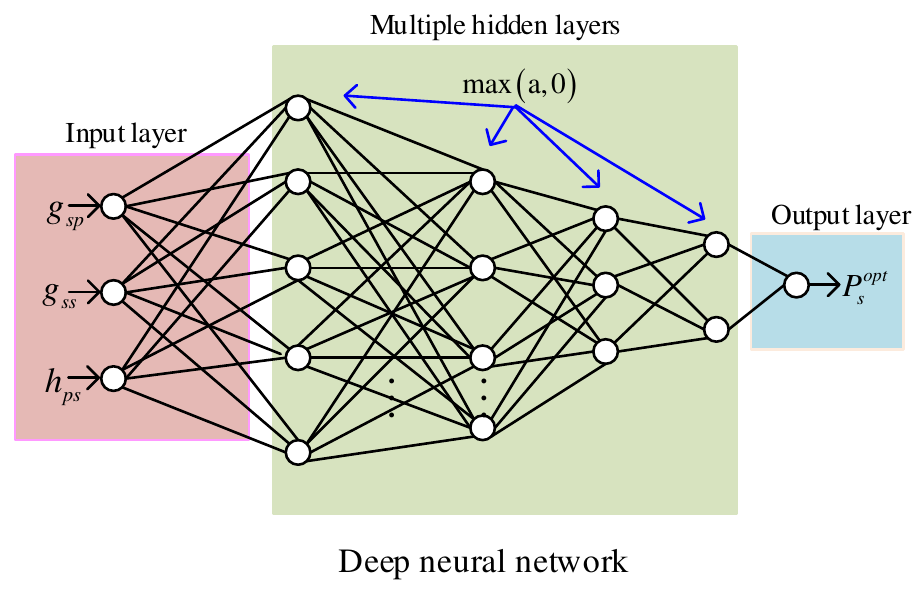}
\caption{The resource allocation framework based on DNN.} \label{fig.1}
\end{figure}

Thus, according to Lemma 3, we propose a resource allocation framework based on DNN to optimize the performance of CRNs, irrespective of the SE and EE of the CRNs, which is shown in Fig. 2. It consists of three layers, namely, the input layer, multiple hidden layers, and the output layer. The inputs are the instantaneous channel power gains $g_{ss}$, $g_{sp}$, and $h_{ps}$, which have continuous probability density functions. The output is the resource allocation strategy of the CBS. In this paper, the output is the optimal transmit power for maximizing the SE of the CRNs  $P_{se}^{opt}$ or the energy-efficient optimal transmit power for maximizing the EE of the CRNs  $P_{ee}^{opt}$.  The activation function for the hidden layers and the output layer is ReLU, namely, $y= {\max } \left(0,x\right)$, where $x$ and $y$ denote the input and output of the neural unit, respectively. The detailed parameters for the network structure is presented in the simulation results.

\begin{figure}[!t]
\centering
\includegraphics[width=3.6 in]{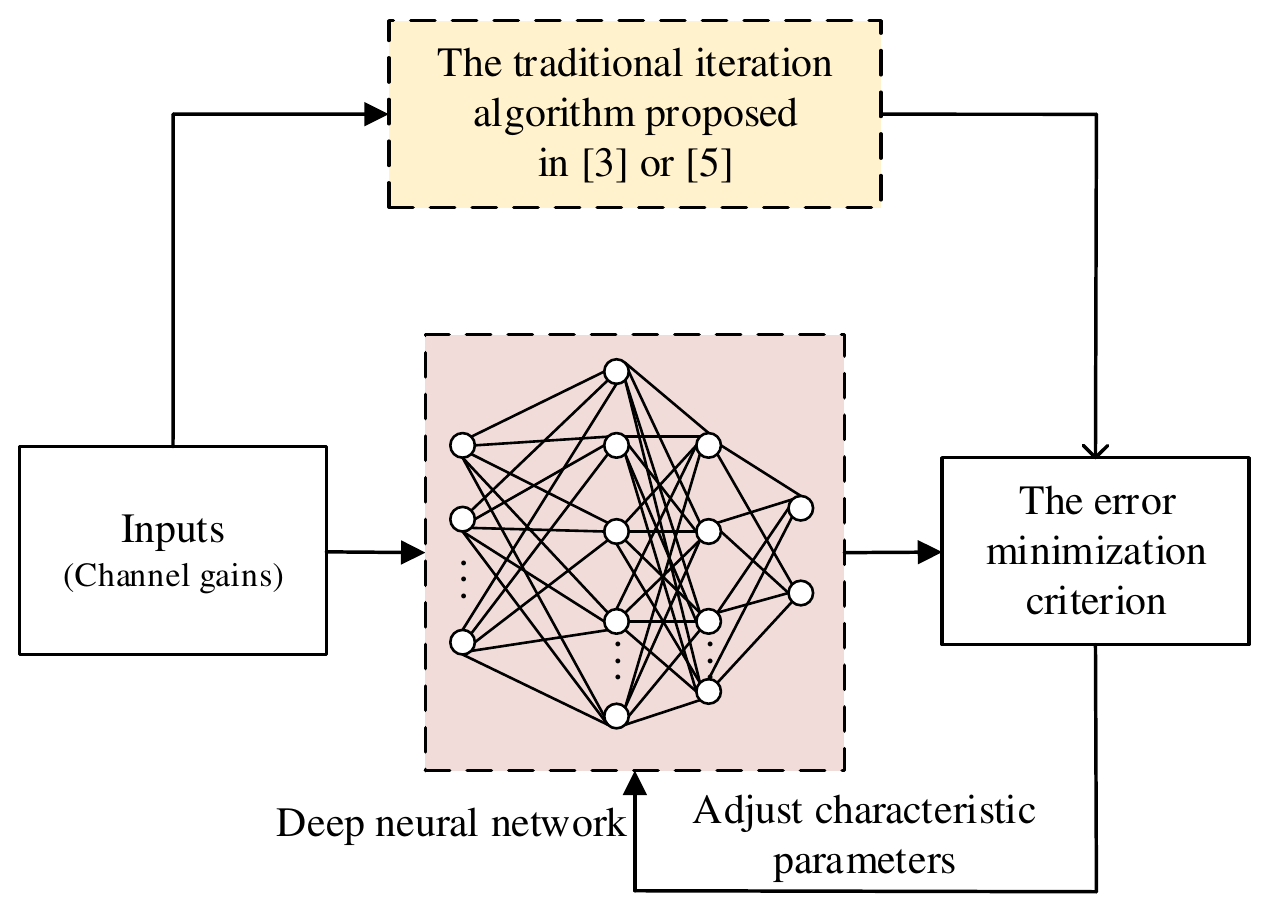}
\caption{The training process for the proposed resource allocation framework based o DNN.} \label{fig.1}
\end{figure}
In order to obtain the weights of each neural unit, the DNN needs to be trained. The training process for the proposed resource allocation framework based on DNN is given in Fig. 3. The training data are obtained by using the conventional resource allocation strategy proposed in \cite{X. Kang} or the energy-efficient resource allocation strategy proposed in \cite{F. H. Zhou1}. The instantaneous channel power gains $g_{ss}$, $g_{sp}$, and $h_{ps}$ have continuous probability density functions, such as exponential distributions. Let $\mathbf{x}_i=[g_{ss}^i \ g_{sp}^i \ h_{ps}^i]$ denote the $i$th input vector of the training process. The output data are the optimal power levels for maximizing the SE obtained from the strategy proposed in \cite{X. Kang} or the energy-efficient optimal power levels obtained from the conventional scheme presented in \cite{F. H. Zhou1}, denoted by $P_{se}^{opt,i}$ and $P_{ee}^{opt,i}$, respectively. Using the conventional scheme, a set with a large volume of training data can be obtained. In the training process, the mean squared error minimization  criterion is applied  \cite{S. Ruder}.  The  mini-batch gradient descent algorithm is exploited to update the weight values since it can be implemented in a distributed manner.

\subsection{Algorithm for Training DNN }
The channel gains are identified as the input and the corresponding optimal transmit power is the output of the DNN  for maximizing the SE or the EE of the CRNs. In this paper, the training process is based on the mini-batch gradient descent algorithm. The training data are firstly divided into $I$ batches and the size of each batch is $M$. For each batch, DNN updates its weights $\theta$ by minimizing the loss function $J(\theta)=\frac{1}{2M} \sum\limits_{m=1}^{M} {(y_m-P_{s,m}^{opt})^2}$, where $y_m$ and $P_{s,m}^{opt}$ is the $m$th output of DNN and the optimal transmit power obtained by using the conventional scheme in a batch, respectively.
Using the gradient descent method, the weights $\theta$ is updated by $\theta^{'}=\theta-\alpha \frac{\partial J(\theta)}{\partial \theta}$, where $\alpha$ is the learning rate of DNN.

The details for the training algorithm is given in Table 1. By utilizing the DNN-based framework, resource allocation can reduce its implementation complexity and achieve real-time performance. It is formally presented as Algorithm 1.

\begin{table}[htbp]
\caption{\label{tab:test}The DNN-based Framework}
\begin{center}
\begin{tabular}{lcl}
\\\toprule
$\textbf{Algorithm 1}$: The DNN-based Framework\\ \midrule
1. Obtain the training data: the channel gains vector $\mathbf{x}[g_{ss} \ g_{sp} \ h_{ps}]$ \\
\ \ \ and the corresponding optimal transmit power $P_{s}^{opt}$ are obtained  \\
\ \ \ by the conventional schemes proposed in \cite{F. H. Zhou1} and \cite{X. Kang};\\
2. Divide the training data into $I$ batches, the size of each batch  is $M$;\\
3. Initialize the weights $\theta$ and the learning rate $\alpha$;\\
4. \textbf{For} each batch $b_i(i=1,2,...,I)$ \textbf{do}\\
\ \ \ \ \ \ Make $\mathbf{x}_i=[g_{ss}^i \ g_{sp}^i \ h_{ps}^i]$ as the $i$th input of the DNN model,\\
\ \ \ \ \ \  obtain the $i$th output $y_i$;\\
\ \ \ \ \ \ Update the DNN's weights $\theta$ by minimizing the loss function \\
\ \ \ \ \ \ $J(\theta_i)=\frac{1}{2M} \sum\limits_{m=1}^{M} {(y_m-P_{s,m}^{opt})^2}$, and $\theta_{i+1}=\theta_{i}-\alpha \frac{\partial J(\theta_i)}{\partial \theta_i}$;\\
\ \ \ \textbf{endfor}\\
5. Save the trained DNN model.

\\\bottomrule
\end{tabular}
\end{center}
\end{table}

\section{Simulation Results}
In this section, simulation results are given to evaluate the performance achieved by using our proposed resource allocation framework. Moreover, the training results are also given to evaluate the training performance. The training process is performed  by using a computation server with four Intel Core i9 CPUs, four Inter Xeon E7-4800 processers,  and 128 GB random access memory. The test results are obtained by using a computer with 8 GB random access memory and Intel Core i7-6500U processor.

The number of neurons in each hidden layer is 200. The training process is based on the data obtained by using the schemes proposed in  \cite{F. H. Zhou1} and \cite{X. Kang} with  $ 1 \times 10^7$ channel realizations. The test results are obtained by using $1 \times 10^3$ channel realizations. The simulation settings are from those in \cite{F. H. Zhou1}. The constant circuit power and  the amplifier coefficient,  $P_C$ and $\zeta$, are set as $0.05 \ W$ and $0.2$, respectively. The variance of the noise is $0.01$. $P_p$ is $60\ mW$. All the iterative step sizes of the subgradient method for updating $\mu$ and $\lambda$ are  $0.1$.  The channels $g_{ss}$, $g_{sp}$, and $h_{ps}$ are all Rayleigh block fading and are  exponential distributed with means $1$, $0.5$, and $0.5$, respectively.

\begin{figure}[!t]
\centering
\includegraphics[width=3.4 in]{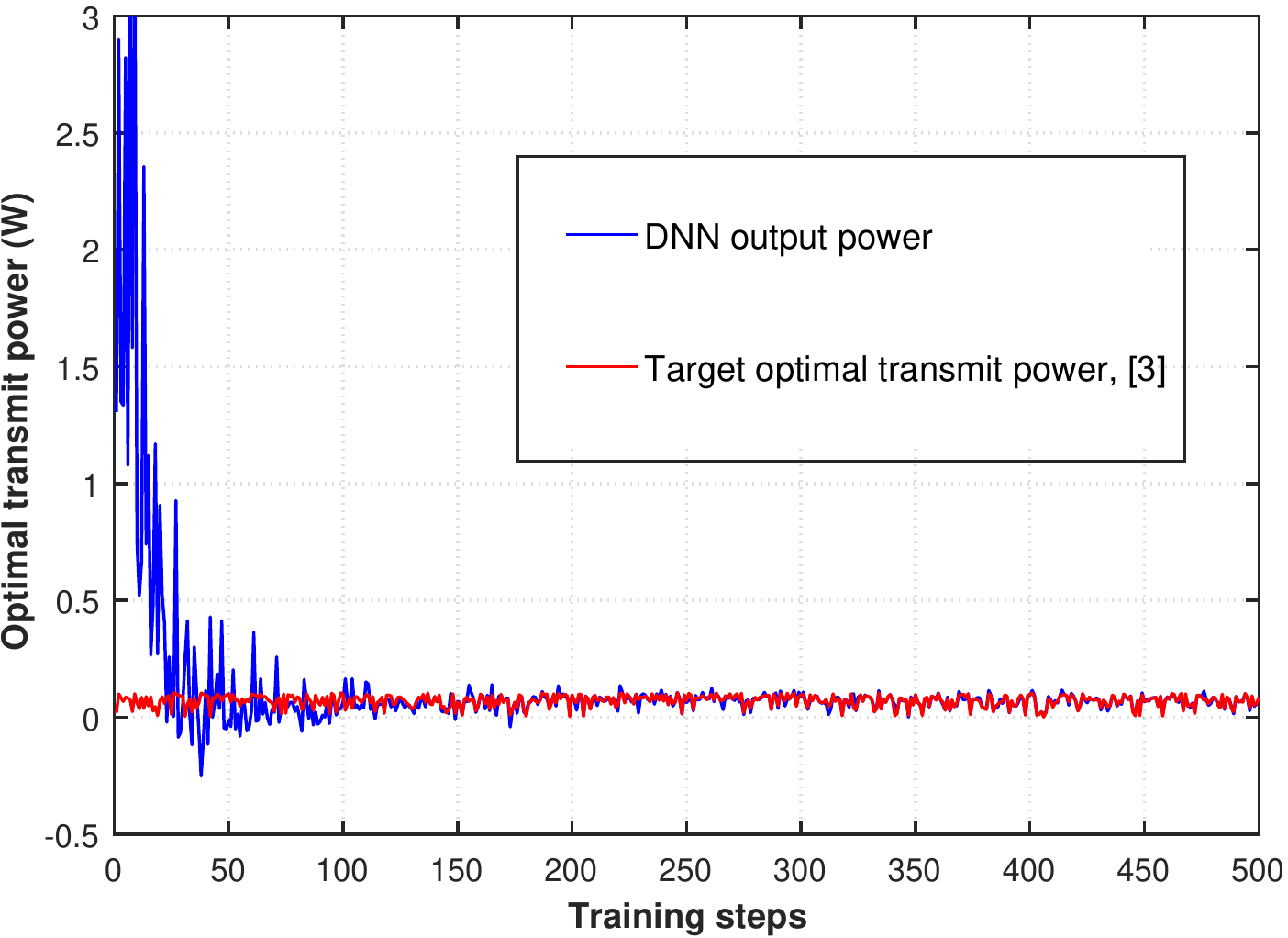}
\caption{The comparison of the optimal transmit power obtained by using the conventional resource allocation strategy \cite{F. H. Zhou1} and achieved by using our proposed resource allocation framework based on DNN at each DNN training step in the training process.} \label{fig.1}
\end{figure}
Fig. 4 shows the comparison of the optimal transmit power obtained by using the conventional resource allocation strategy \cite{F. H. Zhou1} with that achieved by using our proposed resource allocation framework based on DNN at each DNN training step in the training process. For a better presentation, we take samples every 1000 points from the first 500 thousand training steps. It is seen that the outputs of the DNN is increasingly  getting close to the target optimal transmit power with the increase of training steps in the training process. The reason is that the parameters of DNN are continuously updated by using the mini-batch gradient descent algorithm until they are optimal.

\begin{figure}[!t]
\centering
\includegraphics[width=3.4 in]{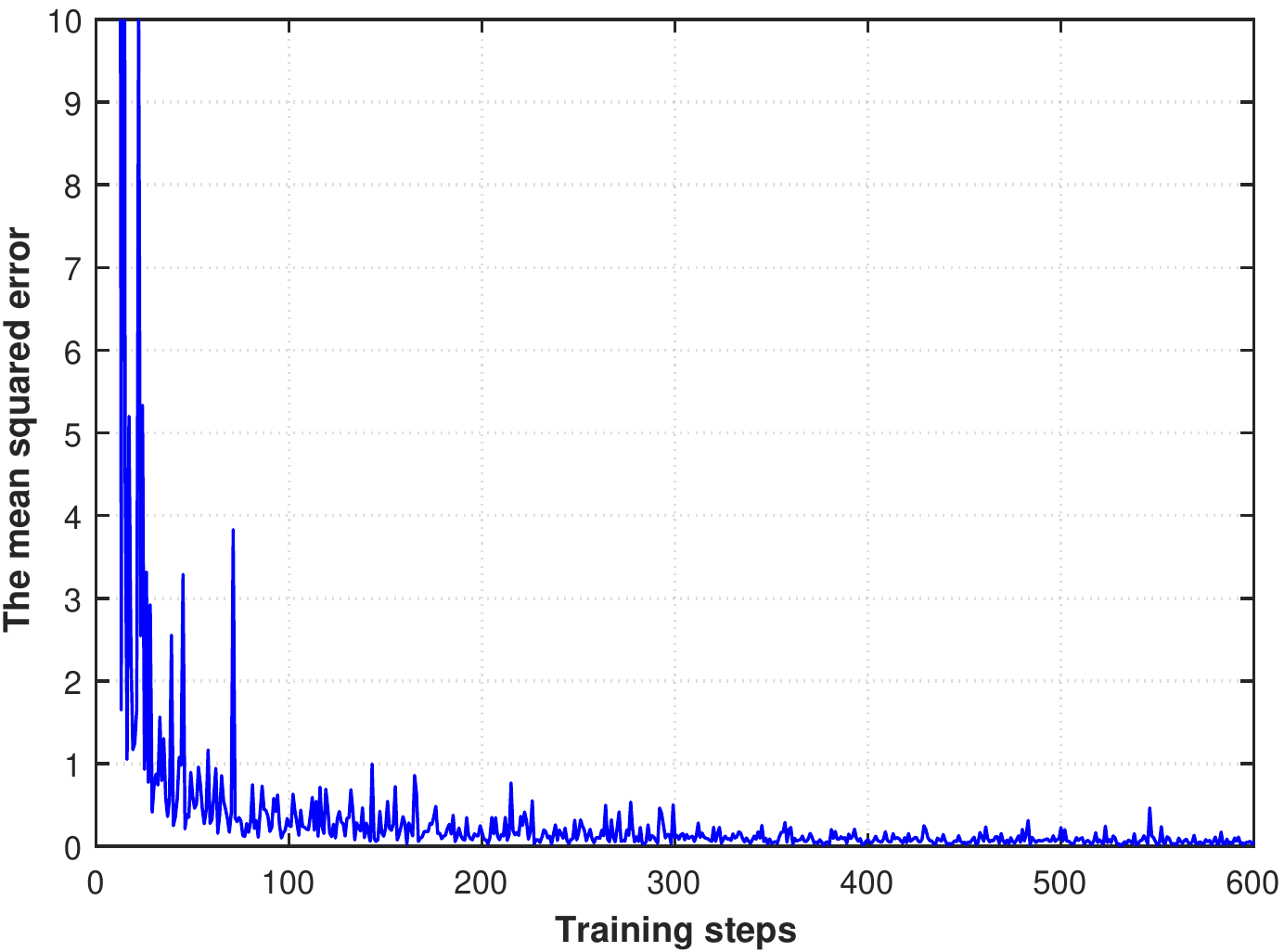}
\caption{The mean square error between the optimal transmit power obtained by using the conventional resource allocation strategy \cite{F. H. Zhou1} and that achieved by using our proposed resource allocation framework based on DNN at each DNN training step in the training process.} \label{fig.1}
\end{figure}
Fig. 5 shows the mean square error between the optimal transmit power obtained by using the conventional resource allocation strategy \cite{F. H. Zhou1} and that achieved by using our proposed resource allocation framework based on DNN at each DNN training step in the training process. The results are obtained by sampling the results from  the first 500 thousand training steps. It can be  seen that the mean square error decreases and approaches  0 with the increase of the training steps in the training process. The reason is similar to that for Fig. 4.

\begin{figure}[!t]
\centering
\includegraphics[width=3.4 in]{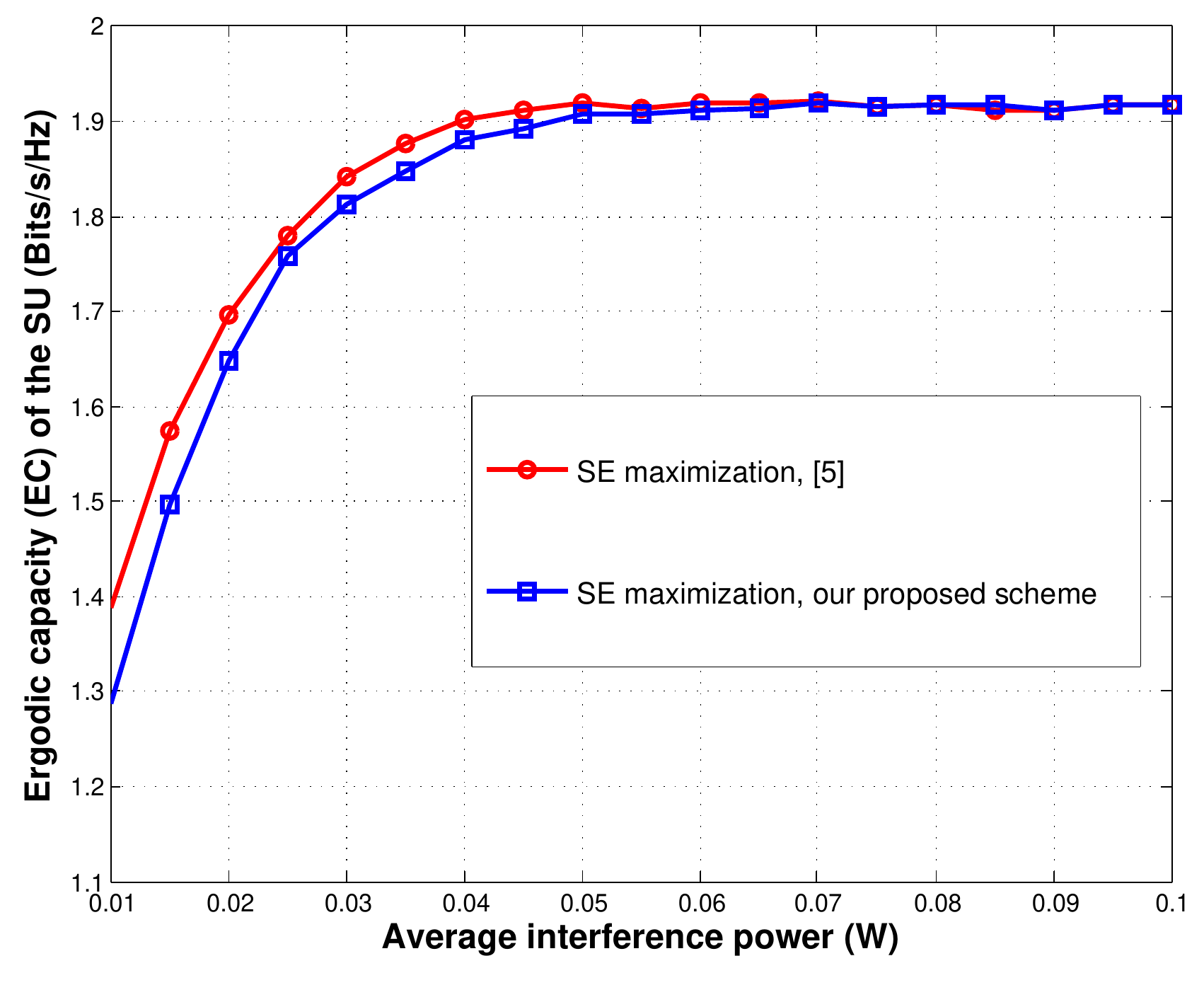}
\caption{The ergodic capacity of the SU versus the AIP constraint for the EE maximization obtained by using the conventional resource allocation strategy and achieved by using our proposed resource allocation framework based on DNN under the PTP/ATP constraint, $P_{th}=100\ mW$.} \label{fig.1}
\end{figure}
Fig. 6 shows the ergodic capacity of the SU versus the AIP constraint for the EE maximization obtained by using the conventional resource allocation strategy and achieved by using our proposed resource allocation framework based on DNN under the PTP/ATP constraint.  The maximum average power of the CBS is set as $P_{th}=100\ mW$. It can be seen that the ergodic capacity achieved by using our proposed scheme is not larger than that obtained by using the conventional scheme proposed in \cite{X. Kang}. The reason is that there is a training error between the output power level and the desired power level. Moreover, it can be seen that the gap between the ergodic capacity achieved by using  our proposed resource allocation framework and that obtained by using the conventional scheme decreases with the average interference power. The reason is that the training error may be decreased when the transmit power of the CBS is relatively large.

\begin{figure}[!t]
\centering
\includegraphics[width=3.4 in]{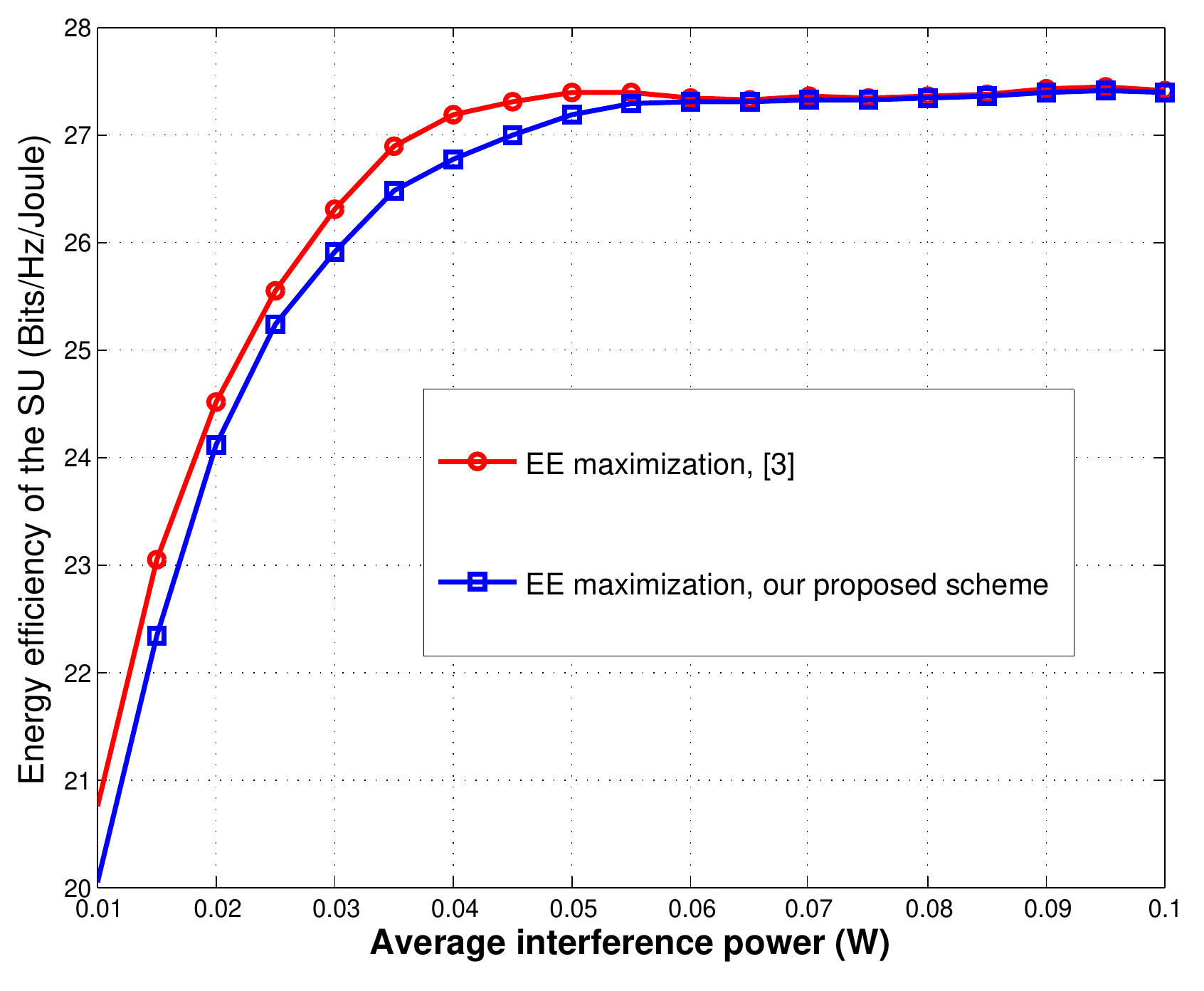}
\caption{The EE of the SU versus the AIP constraint for the EE maximization obtained by using the conventional resource allocation strategy and achieved by using our proposed resource allocation framework based on DNN under the PTP/ATP constraint, $P_{th}=100\ mW$.} \label{fig.1}
\end{figure}
Fig. 7 shows the EE of the SU versus the AIP constraint for the EE maximization obtained by using the conventional resource allocation strategy and achieved by using our proposed resource allocation framework based on DNN under the PTP/ATP constraint. The maximum average power of the CBS is set as $P_{th}=100\ mW$. It is also seen that the EE achieved by using the conventional resource allocation proposed in \cite{F. H. Zhou1} is larger than that achieved by using our proposed scheme. The reason is similar to that for Fig. 4.

\begin{table}[htbp]
 \caption{\label{tab:test}Comparison of the ergodic capacity (Bits/s/Hz) and computation time (sec.)for the SE maximization problem}
 \begin{tabular}{l|c|c|c}
  \midrule
  \midrule
  AIP & Our scheme & Conventional scheme \cite{X. Kang}  & The ratio\\
  \midrule
  \midrule
0.01 W & $1.2874$ & $1.3698$ & $93.98\%$\\
0.02 W & $1.4960$ & $1.5749$  & $94.99\%$\\
0.03 W & $1.6471$ & $1.6985$ & $96.97\%$\\
0.04 W & $1.7590$ & $1.7789$ & $98.88\%$\\
0.05 W & $1.8126$ & $1.8457$ & $98.21\%$\\
0.06 W & $1.8473$ & $1.8764$ & $98.45\%$ \\
\midrule
\midrule
AIP & Our scheme & Conventional scheme \cite{X. Kang}  & The ratio\\
\midrule
\midrule
0.01 W & $0.478$ & $187.53$ & $0.25\%$\\
0.02 W & $0.237$ & $169.48$  & $0.14\%$\\
0.03 W & $0.185$ & $154.84$ & $0.12\%$\\
0.04 W & $0.106$ & $142.79$ & $0.07\%$\\
0.05 W & $0.054$ & $116.42$ & $0.04\%$\\
0.06 W & $0.035$ & $94.73$ & $0.03\%$ \\
\midrule
\midrule
 \end{tabular}
\end{table}
\begin{table}[htbp]
 \caption{\label{tab:test}Comparison of the EE (Bits/Hz/Joule) and computation time (sec.)for the EE maximization problem}
 \begin{tabular}{l|c|c|c}
  \midrule
  \midrule
  AIP & Our scheme & Conventional scheme \cite{F. H. Zhou1}  & The ratio\\
  \midrule
  \midrule
0.01 W & $20.0426$ & $20.7431$ & $96.62\%$\\
0.02 W & $22.3457$ & $23.0420$  & $96.98\%$\\
0.03 W & $24.1126$ & $24.5138$ & $98.36\%$\\
0.04 W & $25.2363$ & $25.5361$ & $98.83\%$\\
0.05 W & $25.8985$ & $26.2969$ & $98.48\%$\\
0.06 W & $26.4802$ & $26.8924$ & $98.47\%$ \\
\midrule
\midrule
AIP & Our scheme & Conventional scheme \cite{F. H. Zhou1}  & The ratio\\
\midrule
\midrule
0.01 W & $4.834$ & $563.52$ & $0.86\%$\\
0.02 W & $3.365$ & $493.73$  & $0.68\%$\\
0.03 W & $2.572$ & $385.77$ & $0.67\%$\\
0.04 W & $1.694$ & $297.95$ & $0.57\%$\\
0.05 W & $1.327$ & $258.38$ & $0.51\%$\\
0.06 W & $0.826$ & $189.79$ & $0.44\%$ \\
\midrule
\midrule
 \end{tabular}
\end{table}
Table II and Table III are given to compare the computation time required by the conventional schemes with that required by using our proposed schemes. It is seen from Table II and Table III that our proposed resource allocation framework based on DNN can significantly decrease the computation time, irrespective of the SE maximization or the EE maximization problem. This verifies the efficiency of our proposed resource allocation framework based on DNN in terms of the computational time and shows the potentials for achieving real-time performance. Moreover, it  is also seen from Table II and Table III that the computational time required for the SE maximization problem is lower than that required for the EE maximization. The reason is that the conventional resource allocation scheme for the EE maximization not only needs to perform iteration, but also requires the sub-gradient method to update the dual variables while the conventional scheme for the SE maximization problem only needs the sub-gradient method.
\section{Conclusions}
In order to achieve a real-time performance and realize  low implementation complexity, the application of DNN into designing resource allocation strategies was studied in CRNs. A resource allocation framework based on DNN was formulated and the training method was presented. Simulation results showed that our proposed resource allocation framework is efficient in terms of the computation time compared with the conventional schemes.


\begin{thebibliography}{20}
\bibitem{F. Zhou4}
Z. Li, L. Guan, C. Li, and A. Radwan, \lq\lq A secure intelligent spectrum control strategy for future THZ mobile heterogeneous networks,\rq\rq \ \emph{IEEE Commun. Mag.}, vol. 56, no. 6, pp. 116-123, June 2018.
\bibitem{Tragos}
E. Tragos, S. Zeadally, A. G. Fragkiadakis, and V. Siris, \lq\lq Spectrum assignment in cognitive radio: A comprehensive survey,\rq\rq \ \emph{IEEE Commun. Surveys Tuts.}, vol. 15, no. 3, pp. 1108-1135, Jul. 2013.
\bibitem{F. H. Zhou1}
F. Zhou, N. C. Beaulieu, Z. Li, J. Si, and P. Qi, \lq\lq Energy-efficient optimal power allocation for fading cognitive radio channels: Ergodic capacity, outage capacity and minimum-rate capacity,\rq\rq \ \emph{IEEE Trans. Wireless Commun.}, vol. 15, no. 4, pp. 2741-2755, Apr. 2016.
\bibitem{F. Zhou3}
F. Zhou, Y. Wu, R. Q. Hu, Y. Wang, and K. K. Wong, \lq\lq Energy-efficient NOMA enabled  heterogeneous cloud radio access networks,\rq\rq \ \emph{IEEE Network}, vol. 32, no. 2, pp.152-160, 2018. 
\bibitem{X. Kang}
X. Kang, Y. C. Liang, A. Nallanathan, H. K. Garg, R. Zhang, \lq\lq Optimal power allocation for fading channels in cognitive radio networks: ergodic capacity and outage capacity,\rq\rq \ \emph{IEEE Trans. Wireless Commun.}, vol. 8, no. 2, pp. 940-950, 2009.
\bibitem{X. Kang1}
F. Zhou, Y. Wu, Y. Liang, Z. Li, Y. Wang, and K. Wong, \lq\lq State of the art, taxonomy, and open issues on NOMA in cognitive radio networks,\rq\rq \ \emph{IEEE Wireless Commun. Mag.}, vol. 25, no. 2, pp.100-108, 2018.
\bibitem{A. Alabbasi}
A. Alabbasi, Z. Rezki, and B. Shihada, \lq\lq Energy efficient resource allocation for cognitive radios: a generalized sensing analysis,\rq\rq \ \emph{IEEE Trans. Wireless Commun.}, vol. 14, no. 5, pp. 2455-2469, May 2015.
\bibitem{X. Zhang}
X. Zhang, H. Li, Y. Lu, and B. Zhou, \lq\lq Distribution energy efficiency optimization for MIMO cognitive radio network,\rq\rq \ \emph{IEEE Commun. Lett.}, vol. 19, no. 5, pp. 847-850, May 2015.
\bibitem{L. Wang}
L. Wang, M. Sheng, X. Wang, Y. Zhang, and X. Ma, \lq\lq Mean energy efficiency maximization in cognitive radio channels with PU outage constraint,\rq\rq \ \emph{IEEE Commun. Lett.}, vol. 19, no. 2, pp. 287-290, Feb. 2015.
\bibitem{D. Wu}
D. Wu, J. Wang, R. Q. Hu, Y. Cai, and L. Zhou, \lq\lq Energy-efficient resource sharing for mobile device-to-device multimedia communications,\rq\rq \ \emph{IEEE Trans. Veh. Technol.}, vol. 63, no. 5, pp. 2093-2103, May 2014.
\bibitem{Q. Li}
Q. C. Li, R. Q. Hu, Y. Xu, and Y. Qian, \lq\lq Optimal fractional frequency reuse and power control in the heterogeneous wireless networks,\rq\rq \ \emph{IEEE Trans. Wireless Commun.}, vol. 12, no. 6, pp. 2658-2668, June, 2013.
\bibitem{A. Yaqot}
A. Yaqot and P. A. Hoeher, \lq\lq Efficient resource allocation in cognitive networks,\rq\rq \ \emph{IEEE Trans. Veh. Technol.}, vol. 66, no. 7, pp. 6349-6361, July 2017.
\bibitem{L. Fan}
L. Wei, R. Q. Hu, Y. Qian, and G. Wu, \lq\lq Energy efficiency and spectrum efficiency of multihop device-to-device communications underlaying cellular networks,\rq\rq \ \emph{IEEE Trans. Veh. Technol.}, vol. 65, no. 1, pp. 367-380, May 2016.
\bibitem{M. Chen1}
M. Chen, W. Saad, C. Yin, and M. Debbah, \lq\lq Echo state networks for proactive caching in cloud-based radio access networks with mobile users,\rq\rq \ \emph{IEEE Trans. Wireless Commun.}, vol. 16, no. 6, pp. 3520-3535, June 2017.
\bibitem{H. Sun}
H. Sun, X. Chen, Q. Shi, M. Hong, X. Fu, and N. D. Sidiropoulos, \lq\lq Training deep neural networks for wireless resource management,\rq\rq \ \emph{IEEE Trans. Signal Process.}, submitted, 2017, https://arxiv.org/abs/1705.09412.
\bibitem{F. Zhou5}
L. Zhou, R. Q. Hu, Y. Qian, and H. H. Chen, \lq\lq Energy-spectrum efficiency tradeoff for video streaming over mobile ad hoc networks,\rq\rq \ \emph{IEEE J. Sel. Areas Commun.}, vol. 31, no. 5, pp. 981-991, May, 2013.
\bibitem{S. Liang}
S. Liang and R. Srikant, \lq\lq Why deep neural networks for function approximation?,\rq\rq \ \emph{ICLR}, 2017.
\bibitem{S. Ruder}
S. Ruder,  \lq\lq An overview of gradient descent optimization algorithms,\rq\rq \ available online: arXiv:1609.04747, Sep. 2016.
\bibitem{M. Chen}
M. Chen, U. Challita, W. Saad, C. Yin, and M. Debbah, \lq\lq Machine learning for wireless networks with artificial intelligence: A tutorial on neural networks\rq\rq \  available online: arxiv.org/abs/1710.02913, Oct. 2017.
\end{thebibliography}
\end{document}